\documentclass[manuscript]{aastex}

\usepackage{xcolor}
\usepackage{graphicx}

\shorttitle{Revisiting the Parker-Moffatt problem}
\shortauthors{Pezzi et al.}

\begin{document}

\title{Revisiting a classic: the Parker-Moffatt problem}

\author{O. Pezzi$^1$, T.N. Parashar$^2$, S. Servidio$^1$, F. Valentini$^1$, C.L. V\'asconez$^3$, Y. Yang$^2$, F. Malara$^1$, W.H.
Matthaeus$^2$ and P. Veltri$^1$}
\affil{
$^1$Dipartimento di Fisica, Universit\`a della Calabria, 87036 Rende (CS), Italy.\\ 
$^2$Department of Physics and Astronomy, University of Delaware, DE 19716, USA. \\
$^3$Departamento de F\'isica, Escuela Polit\'ecnica Nacional, Quito, Ecuador.}

\input epsf 

%\pacs{}

\begin{abstract}
The interaction of two colliding Alfv\'en wave packets is here described by means of magnetohydrodynamics (MHD) and hybrid 
kinetic numerical simulations. The MHD evolution revisits the theoretical insights described by Moffatt, Parker, 
Kraichnan, Chandrasekhar and Els\"asser in which the oppositely propagating large amplitude wave packets interact for a finite 
time, initiating turbulence. However, the extension to include compressive and kinetic effects, while maintaining the gross 
characteristics of the simpler classic formulation, also reveals intriguing features which go beyond the pure MHD treatment.
\end{abstract}


%\date{\today}
%\maketitle

\section{Introduction} 
\label{sec:intro}

A familiar perspective on the hydromagnetic description of astrophysical and laboratory plasma turbulence begins with the 
interaction of oppositely propagating large amplitude incompressible wave packets \citep{iroshnikov64,kraichnan65}. Various 
nonlinear phenomenologies are built on this paradigm \citep{DMV80a,velli89, MatthaeusEA99,galtier00,verdini09,NgBhattacharjee}. 
An essential feature is that large amplitude perturbations in which velocity ${\bf u}$ and magnetic field ${\bf b}$ fluctuations 
are Alfv\'enically correlated, i.e. either ${\bf u} = (c_A/B_0) {\bf b}$ or  ${\bf u} = - (c_A/B_0) {\bf b}$ (where $c_A$ and 
$B_0$ are uniform background Alfv\'en velocity and magnetic field, respectively), are exact stable solutions to the equations of 
incompressible magnetohydrodynamics (MHD) \citep{elsasser50,chandrasekhar56}.  One thread emerging from this concerns the 
analysis of colliding wave packets to reveal properties of the MHD turbulence spectrum \citep{kraichnan65}. 

A different emphasis was given by \citet{moffatt78} and \citet{parker79}. Both of these treatments analyzed the collision of 
large amplitude incompressible, ideal Alfv\'en wave packets noting that nonlinear interaction and mutual distortion of the 
wave packets are limited to the span of time during which they spatially overlap. Both {\it Moffatt} and {\it Parker} argued 
(somewhat differently, as discussed later) that the packets eventually separate and propagate once again undisturbed without 
further interactions. This paper addresses two questions that arise when trying to apply this physical insight to high 
temperature extraterrestrial plasmas such as the solar wind, where such large amplitude Alfv\'enic fluctuations are routinely 
observed \citep{BelcherDavis71}, or solar corona, where the interaction of Alfv\'enic wave packets is thought to occur 
\citep{MatthaeusEA99-coronal}. First, both compressibility effects and kinetic plasma are likely to be important in space 
applications, and we ask if these give rise to significant departures from the the Parker-Moffatt scenario. Second, we ask whether 
the proposed separation of the packets after collision is realized as envisioned, or if a wake of non-propagating disturbances 
might remain after very long times. We address these specific questions using a compressible MHD model and a hybrid Vlasov model.
Broader questions that emerge will be discussed in the concluding remarks. 

In dealing with low Mach number quasi-incompressible fluid or MHD models, either in numerical simulations 
\citep{orszag71,OrszagPatterson72,KraichnanMontgomery80}, applications \citep{DMV80b,MatthaeusEA99-coronal}, or in analytical 
theory \citep{OrszagLesHouches,OughtonEA06}, one routinely deals with two significant properties: first, the dominant quadratic 
couplings are of the form ${\bf k} = {\bf p} + {\bf q}$, transferring energy into (or from) Fourier mode with wave-vector $\bf k$ 
due to nonlinear interactions with  modes at wave-vectors $\bf p$ and $\bf q$. One concludes that in general (unless, e.g., all 
excited wave vectors are co-linear) one expects excitations to spread rapidly among many wave-vectors, a process that over time 
can produce complex mixing and turbulent flows. Second, incompressible MHD nonlinear evolution proceeds as $\partial  z^+_i 
/\partial t \sim -z^-_j \nabla_j z^+_i$ and $\partial  z^-_i /\partial t \sim -z^+_j \nabla_j z^-_i$ in terms of Els\"asser 
variables $z^\pm_j = u_j \pm b_j$ ($j$th components of velocity field $u_j$ and magnetic field $b_j$ in Alfv\'en speed units), 
thus allowing the immediate conclusion that nonlinear couplings vanish unless the Els\"asser fields ${\bf z}^+$ and ${\bf z}^-$ 
have nonzero overlap somewhere in space. These properties not only provide motivation for the Alfv\'en wave packet collision 
problem, but also enter into some of its complexity as an elementary interaction that generates turbulence 
\citep{kraichnan65,DMV80b,howes13,drake16}. 

Beyond the assumption of incompressibility, we may anticipate genuinely compressible and kinetic effects that warrant examination 
in the large amplitude wave packets collision problem. In the solar wind for example, many intervals, especially within 1 AU  
\citep{BrunoEA} or at high latitudes \citep{McComasEA00} are highly Alfv\'enic, but even within such intervals there are mixtures 
of Els\"asser amplitudes, small density variations, and a small parallel variance, as in the well-quoted ``5:4:1'' variance ratio 
reported by \citet{BelcherDavis71}. There have also been reports of interplanetary magneto-sonic wave packets interaction 
\citep{he15}, while the great power-law in the interstellar medium \citep{ArmstrongEA81} is associated with electron density 
fluctuations that may be either propagating or non-propagating \citep{ZankMatthaeus92}. Furthermore in plasmas such as the solar 
wind, at smaller scales near the ion inertial scale, one expects kinetic properties \citep{AlexandrovaEA08} such as spectral 
steepening \citep{bruno13}, dispersive wave effects \citep{SahraouiEA07,GaryEA10} of both Kinetic Alfv\'en Wave and whistler 
types, along with temperature anisotropy, beams and other distortions of the proton velocity distribution function (VDF)
\citep{marsch06,valentini07,servidio12,valentini14,servidio15}. These complications place the problem of collisions of Alfv\'en 
wave packets in a much more complex framework. 

\section{Models and Approach} 
\label{sec:model}

Motivated by these considerations, we revisit the problem of two colliding large amplitude Alfv\'en wave packets by means of 
compressible fluid and kinetic Vlasov-Maxwell simulations. The shown results are from a magnetohydrodynamics (MHD) model and a 
hybrid kinetic plasma (HVM) model. Both retain $2.5$ dimensions in the physical space, with three Cartesian fluid velocities and 
field components, but with gradients only in the $(x,y)$ plane. The HVM model has also a three dimensional velocity space grid.

It is clear that this problem would be better addressed in a high resolution fully 3D representation in physical space. However, 
computational cost limits accessible 3D simulations to relatively low spatial resolution. This present geometry is favored because 
it allows for a large system size, that, in turn, ensures a large Reynolds number and hence MHD like turbulent dynamics 
\citep{parasharEA15}, as well as a realistic realization of compressive fluctuations/parallel variances  \citep{ParasharEA16}.

The dimensionless MHD equations are:
\begin{eqnarray}
& \partial_t \rho +\nabla \cdot (\rho {\bf u})=0 \label{eq:HMHD1} \\ 
& \partial_t {\bf u} +({\bf u}\cdot \nabla){\bf u}= -\frac{{\tilde \beta}}{2\rho}\nabla (\rho T)+ \frac{1}{\rho}
\left[(\nabla \times {\bf B})\times {\bf B}\right]\label{eq:HMHD2} \\
& \partial_t {\bf B} = \nabla \times \left( {\bf u}\times {\bf B} \right) \label{eq:HMHD3} \\
& \partial_t T + ({\bf u}\cdot \nabla)T + (\gamma -1)T(\nabla \cdot {\bf u})=0 \label{eq:HMHD4}
\end{eqnarray}
In Eqs. (\ref{eq:HMHD1})--(\ref{eq:HMHD4}) spatial coordinates ${\bf x}=(x,y)$ and time $t$ are respectively normalized to
$\tilde{L}$ and $\tilde{t}_A=\tilde{L}/\tilde{c}_A$. The magnetic field ${\bf B}={\bf B}_0 + {\bf b}$ is scaled to the typical
magnetic field ${\tilde B}$, while mass density $\rho$, fluid velocity ${\bf u}$, temperature $T$ and pressure $p=\rho T$ are
scaled to typical values ${\tilde \rho}$, ${\tilde c}_A={\tilde B}/(4\pi {\tilde \rho})^{1/2}$, ${\tilde T}$ and ${\tilde
p}=2\kappa_B {\tilde \rho}{\tilde T}/m_p$ (being $\kappa_B$ the Boltzmann constant and $m_p$ the proton mass), respectively.
Moreover, ${\tilde \beta}={\tilde p}/({\tilde B}^2/8\pi)$ is a typical value for the kinetic to magnetic pressure ratio; $\gamma
=5/3$ is the adiabatic index. Details about the numerical algorithm can be found in \citet{vasconez15,pucci16}.

The kinetic simulations solve the hybrid Vlasov-Maxwell equations system \citep{valentini07} in which the proton distribution 
function is numerically evolved while electrons are a massless Maxwellian, isothermal fluid. Dimensionless HVM equations are:
\begin{eqnarray}
& \partial_t f + {\bf v} \cdot \nabla f + \frac{1}{{\tilde \epsilon}} \left( {\bf E}+{\bf v}\times{\bf 
B}\right)\cdot \nabla_{\bf v} f =0 \label{eq:hvm1} \\ 
& {\bf E}= -{\bf u}\times {\bf B} + \frac{{\tilde \epsilon}}{n} \left({\bf j}\times{\bf B}-\frac{{\tilde \beta}}{2}\nabla
P_e\right) \label{eq:hvm2} \\
& \frac{\partial {\bf B}}{\partial t}=-\nabla \times {\bf E} \;\;\; ; \;\;\; \nabla \times {\bf B}={\bf j} \label{eq:hvm3}
\end{eqnarray}
where $f = f({\bf x},{\bf v},t)$ is the proton distribution function. In Eqs. (\ref{eq:hvm1})--(\ref{eq:hvm3}), velocities ${\bf
v}$ are scaled to the Alfv\'en speed ${\tilde c_A}$, while the proton number density $n=\int f\, d^3v$ and the proton bulk 
velocity ${\bf u}=n^{-1}\int {\bf v} f\, d^3v$, obtained as moments of the distribution function, are normalized to 
$\tilde{n}=\tilde{\rho}/m_p$ and $\tilde{c}_A$, respectively. The electric field $\bf{E}$, the current density ${\bf j}= {\bf 
\nabla}\times{\bf B}$ and the electron pressure $P_e$ are scaled to $\tilde{E}=(\tilde{c}_A \tilde{B})/c$, 
$\tilde{j}=c\tilde{B}/(4\pi\tilde{d_p})$ and $\tilde{p}$, respectively. Finally, the parameter ${\tilde \epsilon} = {\tilde 
d}_p/{\tilde L}$, where ${\tilde d}_p={\tilde c}_A/{\tilde \Omega}_{cp}$ is the proton skin depth, allows a comparison of fluid 
and kinetic scales. Electron inertia effects have been neglected in the Ohm's law and no external resistivity $\eta$ is 
introduced. A detailed description of the HVM algorithm can be found in \citet{valentini07,vasconez15,servidio15}.

The simulations are set up as follows. The spatial domain $D(x,y)=[0,8\pi]\times[0,2\pi]$ has been discretized with 
$(N_x,N_y)=(1024,256)$ mesh points, thus implying $\Delta x =\Delta y$ and also an anisotropic wave-vectors space. Spatial
boundary conditions are periodic. In the HVM run, we adopted a uniform velocity space grid with $51$ points in each direction, in 
the region $v_i=[-v_{max},v_{max}]$, being $v_{max}= 2.5{\tilde c}_A$. Velocity domain boundary conditions assume $f=0$ for $|v_i| 
> v_{max}$ ($i=x,y,z$). In our simulation we set $\beta_p=2v_{th,p}^2/{\tilde c}_A^2={\tilde \beta}/2=0.5$, which corresponds to 
$v_{max}=5v_{th,p}$, and ${\tilde \epsilon} =9.8\times10^{-2}$. The background magnetic field is mainly perpendicular to the $x-y$ 
plane, indeed $\mathbf{B_0} = B_0(\sin \vartheta, 0 , \cos \vartheta)$, being $\vartheta=\cos^{-1} \left[\left(\mathbf{B_0} \cdot 
{\hat{\bf z}} \right)/B_0\right] = 6^\circ$ and $B_0=|\mathbf{B_0}|$. 

We impose large amplitude magnetic ${\bf b}$ and bulk velocity ${\bf u}$ perturbations. Density perturbations are not imposed, 
which implies nonzero total pressure fluctuations. In the HVM case, the proton VDF is initially Maxwellian at each spatial 
point. Initial perturbations consist of two Alfv\'enic wave packets with opposite velocity-magnetic field correlation, separated 
along $x$. Because $B_{0,x}\neq 0$, the packets counter-propagate. The nominal time for the collision, which has been evaluated 
from the center of each wave packet, is about $\tau\simeq58.9$. 

The magnetic field perturbation $\mathbf{b}$ has been created by initializing energy in the first four wave-numbers in the $y$ 
direction while, due to the $x$ spatial localization (enforced by projection), many wave-numbers along $x$ are excited initially.
Then, a small $b_z(x,y)$ component has been introduced in such a way that the transverse condition, $\mathbf{B_0} \cdot \mathbf{b} 
= 0$, is hold in each domain point. Finally, the velocity field perturbation $\mathbf{u}$ is generated by imposing that 
$\mathbf{u}$ and $\mathbf{b}$ are correlated (anti-correlated) for the wave packet which moves against (along) the magnetic field 
$B_{0x}$. The intensity of the perturbation is $\langle b\rangle_{rms}/B_0 =0.2$, therefore the Mach number is $M_s = \langle 
u\rangle_{rms}/ v_{th,p} = 0.4$. The intensity of fluctuations with respect to the in-plane field $B_{0x}$ is quite strong, 
with a value of about $2$. It is worth to note that the inverse of the intensity of the fluctuations with respect to the 
in-plane magnetic field is related to the parameter $\tau_{NL}/ \tau_{coll}$, where $\tau_{NL}$ is the characteristic nonlinear 
time and $\tau_{coll}$ is the characteristic collision time. If $\tau_{NL}/ \tau_{coll} \lesssim 1$, several nonlinear times occur 
in a single collision and wave packets can be significantly perturbed by nonlinear effects. On the other hand, if $\tau_{NL}/ 
\tau_{coll} > 1$, many collisions are necessary to strongly distort wave packets. By evaluating $\tau_{NL} \simeq \Delta/u$ (wave 
packet width $\Delta$, perturbations amplitude $u$) and $\tau_{coll} \simeq \Delta/V$ (in-plane Alfv\'en propagation speed 
$V\simeq 0.1 c_A$), it turns out that $\tau_{NL}/ \tau_{coll} \simeq 0.5$. Therefore our simulations stand in a parameter range 
where nonlinear effects can be such important that a strong turbulence scenario may be present.

\subsection{Discussion of the Initial Conditions} 
\label{sec:IC}

The imposed initial perturbations correspond to two large amplitude Alfv\'en wave packets in the sense that magnetic and velocity 
perturbations are fully correlated in each packet, and the packets are separated in space. With zero density variation, a weak 
in-plane uniform magnetic field, and a relatively strong out of plane uniform magnetic field, this initial condition is one 
for which the reasoning of Moffatt and Parker discussed above would be applicable in the context of an incompressible model. 

In addition, the initial data also exactly satisfy the transversality condition $\mathbf{B_0} \cdot \mathbf{b} = 0$, which in 
linear compressible MHD would correspond to the Alfv\'en eigenmode, if indeed the amplitude were infinitesimal. Here the 
amplitude is large, so small amplitude theory is unlikely to be relevant to the nonlinear evolution. Furthermore, the condition
of the proper Alfv\'en eigenmode obtained in large amplitude compressible MHD theory, namely $B=|{\bf B}|= const$ is not 
satisfied by our initial perturbations \citep{Barnes79}. This suggests that pressure and density fluctuations may be generated 
during the wave packets evolution. Therefore the initial data are nonlinear eigenmodes of incompressible MHD, but not exact 
eigenmodes of compressible MHD. On the other hand we do not expect significant differences because the initial $B=|{\bf B}|$ 
fluctuations are not very large (less than $10\%$). Future works will analyze the evolution in the case in which $B= const$ at 
$t=0$, and in the framework of other plasma models, such as Hall MHD and Particle In Cell (PIC) kinetic plasma models.

\section{Numerical Results} 
\label{sec:results}

\begin{figure*}[!htb]
\epsfxsize=15cm \centerline{\epsffile{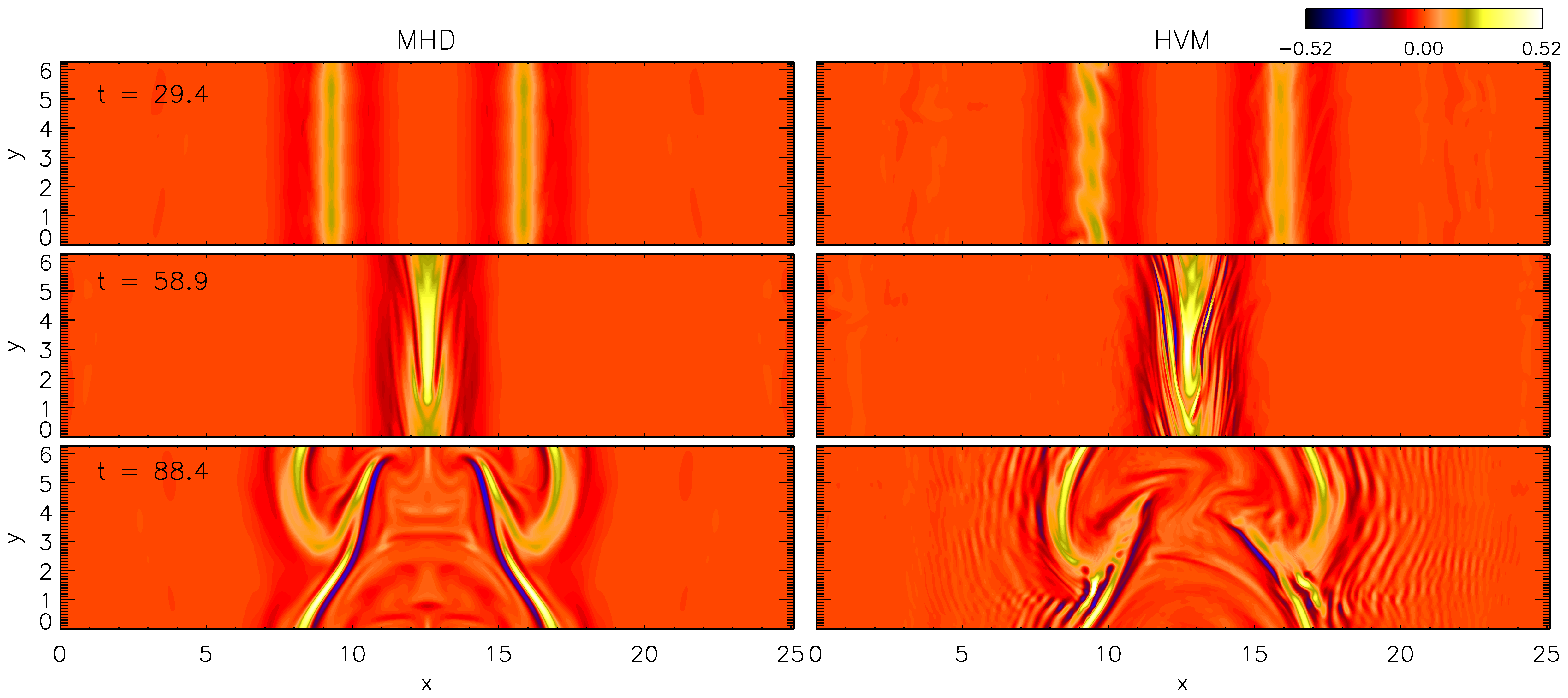}}   % FIGURE N.1
\caption{(Color online) Contour plots of the current density $j_z(x,y)$ for the MHD (left column panels) and the HVM (right
column panels) simulations. Top, center and bottom rows correspond respectively to the time instants $t=29.4$, $t=\tau=58.9 $ and
$t=88.4$.}
\label{fig:jz}
\end{figure*}

An overview of the dynamics in the two simulations (MHD and HVM) can be appreciated by inspection of the evolution of the 
out-of-plane component of the current density $j_z(x,y)$, reported in Fig. \ref{fig:jz}. Left and right columns of Fig. 
\ref{fig:jz} refer to the MHD and HVM simulations, respectively. The rows refer to different instants of time: top, center and 
bottom rows respectively indicate $t=29.4$, $t=\tau=58.9$ and $t=88.4$. In both simulations the initially separated wave packets 
counter-propagate, approach each other [top panels of Fig. \ref{fig:jz}], and collide at $t=\tau$. During the collision [center 
panels of  Fig. \ref{fig:jz}], $j_z$ intensifies, and, since the overlapping wave packets interact nonlinearly, the dynamics 
produces small scales that can be easily appreciated by examining the width of the current structures in the center row of Fig. 
\ref{fig:jz}. At the final stages of the simulations [bottom panels of Fig. \ref{fig:jz}], the wave packets continue their motion 
while displaying a significantly perturbed shape. Indeed the $j_z$ contours indicate that current structures are much more complex 
after that the collision occurs. Moreover, their shape exhibits also a curvature which is not anticipated prior to the collision 
and which indicates the presence of energy in modes with gradients along the $y$ direction, transverse to the propagation. 

The MHD and HVM evolutions exhibit noticeable differences: the MHD case is symmetric with respect to $x=L_x/2 \approx 12.5$, 
while this symmetry is lost in the Vlasov run also before the collision [right top panel of Fig. \ref{fig:jz}]. The lack of the 
symmetry in the HVM run may be due to the inter-coupling of $B_{0x}$ and dispersive effects which are present in the HVM run and 
may cause a different propagation along and against $B_{0x}$. The HVM run also forms smaller scales during the interaction than 
in the MHD case. This difference can be appreciated in the right, center panel of Fig. \ref{fig:jz}. Finally, after the collision, 
$j_z$ is much more complex in the HVM simulation than in the MHD run. Very thin current sheet structures and secondary ripples are 
observed only in the Vlasov case. The nature of these secondary ripples may be associated with the presence of some KAW-like 
fluctuations \citep{hollweg99,vasconez15} and will be analyzed in detail in a separate paper. 

\begin{figure}[!b]
\epsfxsize=7.5cm \centerline{\epsffile{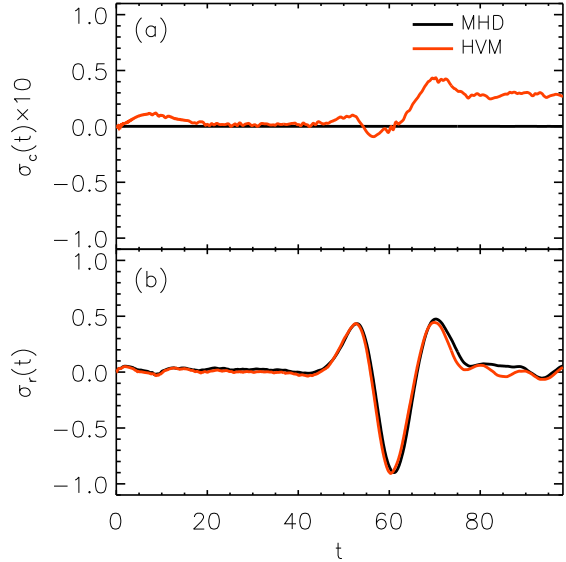}}   % FIGURE N.2
\caption{(Color online) Temporal evolution of $\sigma_c(t)$ (a) and $\sigma_r(t)$ (b). In each panel black and red lines refer
to the MHD and HVM cases, respectively.}
\label{fig:hc}
\end{figure}

A point of comparison of our simulations with respect to the theoretical ideas given by Moffatt and Parker is to examine the 
behavior of cross helicity. Those theoretical treatments assume ideal non-dissipative conditions, so that the total cross 
helicity is conserved and moreover the expectation is that the separate wave packets after the collision have the same cross 
helicity as prior to the interaction. Furthermore the initial and final states, in the ideal treatment, have equipartition of flow 
and magnetic field energy, with departures from equipartition possible during the interaction. To examine these, Fig. 
\ref{fig:hc} shows the temporal evolution of (a) the normalized cross-helicity $\sigma_c(t)$, and (b) the normalized residual
energy $\sigma_r (t)$ \citep{bruno13}, respectively defined as $\sigma_c= (e^+ - e^-) / (e^+ + e^-)$ and $\sigma_r= (e^u - e^b) / 
(e^u + e^b)$, where $e^\pm=\langle(\mathbf{z^\pm})^2\rangle/2$, $e^u=\langle\mathbf{u}^2\rangle/2$, 
$e^b=\langle\mathbf{b}^2\rangle/2$ and $\mathbf{z^\pm}=\mathbf{u}\pm\mathbf{b}$. 

The normalized cross-helicity $\sigma_c(t)$ is well-preserved in the MHD simulation, while, in the kinetic simulation, 
$\sigma_c(t)$ displays, around the collision time $\tau$, a significant growth followed by a saturation stage. It is worth to note 
that, in the MHD run, $\sigma_c(t)$ is still conserved despite $\sigma_c(t)$ is an invariant of the incompressible MHD while our 
MHD simulation is compressible [compressibility could, in principle, break the invariance of $\sigma_c(t)$]. This characteristic 
reflects the fact that the compressive fluctuations which are dynamically generated during the MHD evolution are not strong enough 
to break the invariance of $\sigma_c(t)$. On the other hand, the breaking of the $\sigma_c$ invariance in the Vlasov run is 
associated with the presence of both dispersive and kinetic effects. Indeed, by evaluating the incompressible Hall MHD invariant 
$\sigma_g$ (generalized helicity) \citep{turner86,servidio08}, $\sigma_g$ is also not preserved for the HVM run and it shows a 
similar behavior of $\sigma_c$ (in particular a jump is recovered at $t\simeq\tau$). This feature suggests that the production of 
$\sigma_c(t)$ recovered in the kinetic simulation cannot be associated only with dispersive effects, which are taken into account 
in the Hall model, but is also due to the presence of kinetic effects.

In contrast the evolution of the residual energy $\sigma_r (t)$ is very similar in the MHD and HVM simulations. Referring to 
Fig. \ref{fig:hc} (b), we see that in the initial stages $\sigma_r\simeq 0$, then $\sigma_r$ strongly oscillates during the wave 
packets collisions, first to positive values indicating a positive correlation of the Els\"asser fields, then moving more 
strongly towards negative values of correlation, and returning to positive correlation again prior to finally approaching zero 
once again. It is clear that during the collision there is substantial exchange of kinetic and magnetic energy, and this is
not greatly influenced by kinetic effects. 

\begin{figure}[!tb]
\epsfxsize=6.5cm \centerline{\epsffile{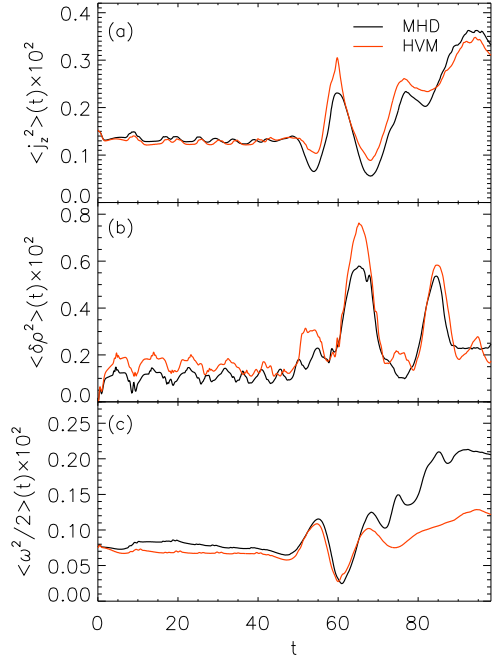}}   % FIGURE N.2
\caption{(Color online) Temporal evolution of $\langle j_z^2\rangle (t) $ (a), and  $\langle \delta \rho^2\rangle (t)$ (b) and 
$\langle\mathbf{\omega}^2/2\rangle (t)$ (c) for the MHD (black line) and HVM (red line) simulations.}
\label{fig:enst}
\end{figure}

In order to compare the role of small scale dynamics in the two cases, we computed the mean square out-of-plane electric current 
density $\langle j_z^2\rangle $ as a function of time. This is illustrated in the top panel of Figure \ref{fig:enst} for the 
MHD (black) and kinetic HVM (red) runs. In both cases, $\langle j_z^2\rangle (t)$ shows a similar time evolution. In particular, 
both models show a peak of $\langle j_z^2\rangle (t)$ around the collision time $t\simeq \tau$. After the collision, a 
high-intensity current activity persists and the peak of current activity is reached in the final stage of the simulations.

Other quantities that provide further physical details about the simulations are $\langle \delta \rho^2\rangle $, the density 
fluctuations providing a measure of compressibility, and $\langle \mathbf{\omega}^2/2\rangle$, the mean square vorticity or 
enstrophy, where $\delta \rho=\rho-\langle\rho\rangle $ and $\mathbf{\omega}={\bf \nabla}\times {\bf u}$. Panels (b) and (c) of 
Figure \ref{fig:enst} respectively show $\langle \delta \rho^2\rangle $ and $\langle \mathbf{\omega}^2/2\rangle$ for both 
simulations. The density fluctuations peak around $t\simeq 65$ and $t\simeq 85$. The first peak is due to the interaction 
between the two wave packets. The second peak of density fluctuations appears to be due to propagation of magneto-sonic 
fluctuations generated by the initial strong collision. Once generated these modes propagate across the periodic box and provide 
an ``echo'' of the original collision. We also note that in the initial stage of both simulations, $\langle \delta \rho^2\rangle$ 
exhibits some small modulations, which could be produced by the absence of a pressure balance in the initial condition.

The enstrophy $\langle \mathbf{\omega}^2/2\rangle $ evolution, also shown in Fig. \ref{fig:enst}, indicates that both MHD and HVM 
cases produce fine scale structure in the velocity, i.e., vortical structures, during the collisions, and these persist after the 
collision. However, in the MHD case  $\langle \mathbf{\omega}^2/2\rangle $ reaches larger values compared to the HVM case. This 
could be due to the presence of kinetic damping effects which decrease the intensity of $\langle 
\mathbf{\omega}^2/2\rangle$ by transferring energy to the VDF [see. e.g., \citet{DelsartoEA16,Parashar16b}]. It is interesting to 
note that the general profile of enstrophy and mean square current follow similar trends in time. This can be expected as the 
inertial range of turbulence typically provides near-equipartition of velocity and magnetic fluctuation energy, even in fairly 
simple configurations \citep{MatthaeusLamkin86}. However, when examined in more detail, one often finds, as here, that the 
magnetic fluctuations are usually about a factor of two more energetic in the inertial range part of the spectrum, as they are, 
for example in the solar wind \citep{MatthaeusGoldstein82a}. This inequality is here reflected in the fact that $\langle 
j_z^2\rangle > \langle \mathbf{\omega}^2/2\rangle $. 

\begin{figure}[!tb]
\epsfxsize=9cm \centerline{\epsffile{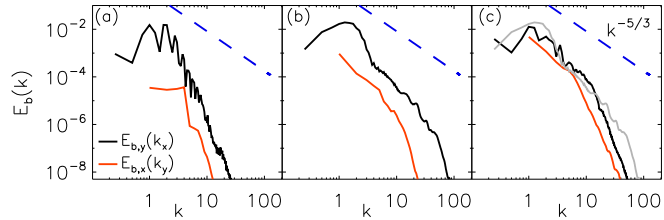}}   % FIGURE N.2
\caption{(Color online) Power spectra density of the magnetic energy $E_b$ at three time instants $t=29.4$ (a), $t=\tau=58.9$ (b)
and $t=88.4$ (c). In each panel the black curve indicates $E_{b,y}(k_x) = \langle E_b(k_x,k_y) \rangle_{k_y}$, the red curve
refers to $E_{b,x}(k_y) = \langle E_b(k_x,k_y) \rangle_{k_x}$ and the dashed blue line indicates the slope $k^{-5/3}$ as a
reference. The gray line in panel (c) represents $E_{b,y}(k_x)$ at $t=58.9$, which corresponds to the black line of panel (b).}
\label{fig:spectra}
\end{figure}

In order to further characterize the small scales fluctuation generated by the collisions and associated nonlinear activity, we 
examine the magnetic energy spectra. Figs. \ref{fig:spectra} shows, for the Vlasov run, the reduced magnetic field power spectral 
densities $E_{b,y}(k_{x})$ (black)  and $E_{b,x}(k_{y})$ (red) at (a) $t=29.4$; (b) $t=\tau=58.9$; and (c) $t=88.4$. Here, for 
example, $E_{b,y}(k_{x})$ is a one-dimensional reduced power spectra density obtained by integrating over one component of 
wave-vector. The blue dashed lines in Fig. \ref{fig:spectra} indicate a $k^ {-5/3}$ slope for reference. We remark that the 
spectra for the HVM run show more energy at very small scales compared to the MHD run (not shown here), again consistent with the 
idea that the HVM run produces more fine scales than the the MHD simulation. This may simply indicate that the effective 
dissipation in the HVM case is smaller than the numerically motivated dissipation coefficients selected for the MHD run. 

In Fig. \ref{fig:spectra} we observe that, at $t=29.4$, the spectrum $E_{b,y}(k_x)$ is steep due to the localization of the 
initial condition, which requires involvement of a wide range of wave-numbers $k_x$. Furthermore, during the evolution, 
the spectra show a transfer of energy towards small scales, at higher $k_x$ and at higher $k_y$. This represents a signature of 
energy transfer due to nonlinear coupling. In fact, much of the energy $E_{b,y}(k_x)$ is contained, at $t=\tau$, in a bump 
around $k=1$, while at $t=88.4$ the bump is less clear and the spectrum $E_{b,y}(k_x)$ is more developed compared to the one at 
$t=\tau$. A break in $E_{b,y}(k_x)$ can be also appreciated around $k_{d_p}\simeq 10$. 

Another feature of the magnetic energy spectra is that $E_{b,x}(k_y)$, which initially contains less energy than $E_{b,y}(k_x)$, 
experiences a significant increase in power, reaching almost the same amplitude at the later times. This suggests that 
fluctuations become more isotropic, and that energy transfer is efficient in both directions of the wave-vector space. 

\section{Conclusions} 
\label{sec:concl}

We have carried out a comparative study using different plasma simulation methods to examine the dynamical evolution that 
accompanies the interaction or “collision” of two oppositely propagating wave packets. For the classic case of incompressible MHD, 
considered by Moffatt and by Parker, the wave packets are, when considered separately, exact large amplitude solutions of the 
nonlinear equations, and are, therefore, strictly speaking ``waves''. If two such waves, oppositely propagating, become 
overlapping, nonlinear couplings and turbulence can be produced and the packets are deformed. Moffatt and Parker concluded that, 
after the characteristic interaction time, the packets again separate and continue propagating away from one another without 
further nonlinear interactions. Hence one question addressed in the present study is whether such separation after a collision 
actually occurs. A second question is whether departures from incompressible MHD change the dynamics in an appreciable way. To 
that end, here we examined a compressible MHD formulation and a hybrid Vlasov formulation.

Our results show that - when one moves beyond the MHD framework in which the Moffatt \& Parker problem is approached - the 
dynamics become more complex. Here we find that, in both the present cases, the interactions and the structures produced in the 
collision are sufficiently complex that it is difficult to determine whether the wave packets actually attain a full separation 
after the collision. Indeed, we note that very complex current and vorticity structures are produced at small scales in both 
compressible MHD and HVM cases and these fluctuations are indicative of a spread of energy in the wave vectors plane, which is 
almost perpendicular to ${\bf B_0}$. The energy spectra evolve toward isotropy in this plane, although one would expect a degree 
of spectral anisotropy to persist due to the presence of the weak in-plane magnetic field. Furthermore, to the extent that the 
interaction of the packets has a finite lifetime [as envisioned e.g. by \citet{kraichnan65}], any such relaxation would be 
expected to be incomplete in a single interaction time.

In addition, we recall that in the incompressible ideal problem, cross helicity is conserved, so that after the collision in that 
case, the separated wave packets will each contain the same energy that was present in the initial state. However, cross-helicity 
is not preserved in the kinetic case since dispersive (e.g. Hall effect) and dissipative effects are present in the simulation. In 
fact in the HVM case, we observe a significant change in global cross helicity during the interaction.

This preliminary examination of the fate of the Moffatt and Parker conjecture in the context of compressible as well as kinetic 
models has produced a satisfactory, if not complete, picture. In fact the basic physics of the large amplitude Alfv\'en waves 
collisions as envisioned by those authors appears to be upheld in this regime. However, we have neither examined the most general 
case, which would require full 3D simulations nor have we completely analyzed the compressible and kinetic effects in the present 
cases. In particular, not shown here are indications of specific kinetic wave modes and characteristic distortions of the velocity 
distribution function \citep{servidio15,vasconez15} that might be expected. A separate account of these results is in preparation. 
We also note that we have analyzed the present problem employing several other models including hybrid particle in cell, Hall MHD 
and also varying the initial conditions. Such results are of interest in the context of e.g. Turbulent Dissipation Challenge 
\citep{ParasharJPP15} and will be reported at a later time.

\begin{acknowledgments}
Research is supported by NSF AGS-1063439, AGS-1156094 (SHINE), AGS-1460130 (SHINE), and NASA grants NNX14AI63G (Heliophysics 
Grandchallenge Theory), and the Solar Probe Plus science team (ISIS/SWRI subcontract No. D99031L), and Agenzia Spaziale Italiana 
under the contract ASI-INAF 2015-039-R.O ``Missione M4 di ESA: Partecipazione Italiana alla fase di assessment della missione 
THOR''. Kinetic numerical simulations here discussed have been run on the Fermi supercomputer at Cineca (Bologna, Italy), 
within the ISCRA-C project IsC37-COLALFWP (HP10CWCE72). 
\end{acknowledgments}

\clearpage

\end{document}